\documentclass[pdflatex,sn-mathphys-num]{sn-jnl}

\usepackage{epsf,float,epsfig}

\usepackage{amsmath}
\usepackage{amssymb}
\usepackage{natbib}

\begin{document}

\title[Quantum Algorithm for Binary Vector Encoding and Retrieval Utilizing the Permutation Trick]{Quantum Algorithm for Binary Vector Encoding and Retrieval Utilizing the Permutation Trick}

\author*{ \sur{Andreas Wichert}}\email{andreas.wichert@tecnico.ulisboa.pt}
\affil{\orgdiv{Department of Computer Science and Engineering},  \orgname{ INESC-ID  \&  Instituto Superior T\'ecnico University of Lisbon}, \orgaddress{ \city{Porto Salvo}, \postcode{2740-122}, \country{Portugal}}}

\abstract{We present a novel quantum storage algorithm for $k$ binary vectors of dimension $m$ into a superposition of a $m$-qubit quantum state based on a permutation technique. We compare this algorithm to the storage algorithm proposed by Ventura and Martinez. The  permutation technique is simpler and can lead to an additional reduction through the \textit{reduce} algorithm.
To retrieve a binary vector from the superposition of $k$ vectors represented by a $m$-qubit quantum state, we must use a modified version of Grover’s algorithm, as Grover’s algorithm does not function correctly for non-uniform distributions. We introduce the permutation trick that enables an exhaustive search by Grover’s algorithm in $O(\sqrt{k})$ steps for $k$ patterns, independent of $n=2^m$. We compare this trick to the Ventura and Martinez trick, which requires $O(\sqrt{n})$ steps for $k$ patterns. }


\maketitle

\section{Introduction}

Quantum encoding is a process that transforms classical information into quantum states. Basis encoding is the most intuitive method for encoding classical binary vectors into a quantum state and plays a crucial role in many quantum algorithms, such as quantum nearest neighbor and quantum associative memory. Basis encoding maps $k$ $m$-dimensional binary vectors into a superposition of a $m$-qubit quantum state, with the condition that $k < 2^m$.
To retrieve a binary vector from the superposition of a $m$-qubit quantum state, a modified version of Grover’s algorithm is employed, as Grover’s algorithm does not function correctly for non-uniform distributions.
Typically, basis encoding corresponds to the storage phase of information, while the application of Grover’s algorithm to the retrieval phase. These two principles will be demonstrated using examples of quantum nearest neighbor and quantum associative memory.

\subsection{Quantum Nearest Neighbor}

The Nearest Neighbors (NN) classifier operates on the premise that objects represented by feature vectors sharing similar characteristics are likely to belong to the same class, \cite{Mitchell97}.  We assume that the vectors are binary. Consequently, if we represent each example as a feature vector, we can compute the Hamming distance between a novel example and some previously encountered training data to identify which examples are similar to the new one. This information can then be utilized for classification purposes.

Let $DB$ be a database of $k$  objects $pattern_i$ represented by binary vectors of dimension $m$ in which the index $i$ is an explicit key identifying each object,
\[ \{ pattern_i \in DB~ |~ k \in \{1.. k\}\}. \]
 with
\[ pattern_1,  pattern_2, \cdots,  pattern_k. \] 
For a query vector $pattern_{query}$  one vector $pattern_i$ is NN-similar to $pattern_{query}$  according to the Hamming distance function indicating the number of different features between two binary vectors $d_H$ with
  \begin{equation}
pattern_i=\min_j  d_H(pattern_j, pattern_{query} ). 
\end{equation}
Since in quantum computation we cannot simply compare  vectors, the  $\epsilon$-similarity is used.
The $\epsilon$-similarity is defined  for a range queries. For a query vector $pattern_{query}$  all vectors $pattern_j$ are $\epsilon$-similar to $pattern_{query}$  according to the Hamming distance function $d_H$ with
  \begin{equation}
d_H(pattern_j, pattern_{query} ) \leq \epsilon,
\end{equation}
in which $\epsilon$ indicates the number of different features. 
The search algorithm is sensitive to the value of $\epsilon$. If $\epsilon$ is too large, the entire dataset is searched. If $\epsilon$ is set to zero, only one solution is found. Even if the search results in multiple patterns, quantum computation can only indicate one pattern.
In quantum computation, the set $DB$ can be represented by a quantum register of $m$ qubits in superposition. The $\epsilon$-similarity is determined by marking solutions using an oracle function. Patterns to the query are identified using the modified Grover’s amplification algorithm, and one solution is measured. This algorithm is known as the quantum nearest neighbor (QNN) algorithm. The QNN algorithm is related to the quantum associative memory (QuAM), which was proposed by Venture and Martinez in 1988, see \cite{Ventura1988, Ventura2000, tay2010}). Unlike the nearest neighbor algorithm, the QuAM is based on cognitive and biological modeling of information retrieval in the human brain.

\subsection{Quantum Associative Memory}

Associative memory models human memory, as evidenced by research in the fields of psychology and neuroscience, \cite{Palm90,Churchland94,Fuster95,Squire99}. These models incorporate several key abilities, including the ability to correct errors when false information is presented, the ability to complete incomplete information, and the ability to interpolate information when a specific pattern is not currently stored,  \cite{Palm82,Hertz91,Anderson95,Kohonen89}.
Over the years, various associative memory models have been proposed ~\cite{Amari1972}, \cite{Hecht-Nielsen89, Kohonen89}, \cite{Anderson95b,Ballard97}, including the Hopfield model, which represents a recurrent model of associative memory. The Hopfield model (see \cite{Amari1972}, \cite{Hopfield1982}, \cite{Hertz91}) is a dynamical system that evolves until it reaches a stable state.  
In binary vector representation, the presence of a feature is denoted by a `one’ component of the vector, while its absence is indicated by a `zero’ component. After learning, the query vector is presented to the associative memory, and the most similar vector, based on the Hamming distance, is determined by the retrieval rule. 
Typically, associative memory operates in two distinct phases: the learning phase, during which  vectors are stored, and the retrieval phase, in which the most similar results to a query vector are identified.

In the realm of quantum computation, QuAM is a model of an associative memory characterized by an exponential capacity in the number of artificial neurons represented by a quantum register.
Unlike traditional associative memory that store patterns in neurons, QuAM stores them as a superposition in a quantum register. This superposition enables simultaneous evaluation of all stored patterns.

The primary contribution of the QuAM model, as proposed by Ventura and Martinez in 1988, \cite{Ventura1988, Ventura2000}, lies in its specification of the learning phase and retrieval phase by a quantum algorithm. During learning phase, vectors are stored in a superposition of $k$ binary vectors with a dimension of $m$. The number of stored patterns $k$ is smaller as the value $n$, with  $n=2^m$, meaning $k < n$.
It is noteworthy that the method employed in the QuAM model is linear in the number of stored patterns and their dimension, \cite{Schuld2018}. 

Uniform distributions are essential for the functioning of Grover’s algorithms. The superposition of $k$ binary vectors does not constitute a uniform distribution. The non uniform distribution is a fundamental requirement for the operation of the quantum associative memory (QuAM), as proposed by Venture and Martinez. To address this, a modified version of Grover’s search algorithm is employed to identify the answer vector corresponding to a query vector, achieving $O(\sqrt{n})$ Grover's rotations. 

\subsection{Organization}

The paper is organized as follows:
\begin{itemize}
\item We describe the storage algorithm of binary vectors by Ventura and Martinez.
\item We introduce the storage algorithm for binary vectors based on permutation technique.
\item We present the storage algorithm and the \textit{reduce} algorithm with an example.
\item We describe the Ventura Martinez trick for Grover's algorithm.
\item We compare  the Ventura Martinez trick with the permutation trick  for Grover's algorithm.
\end{itemize}
 
\section{Storage for Binary Vectors}

 Classical information, represented as binary vectors, is mapped into a superposition of quantum states through quantum encoding.

A uniform superposition of binary vectors can be effortlessly generated using Hadamard gates. For instance, when employing four qubits, the uniform superposition is achieved through the application of Hadamard gates
\[
 H_4 |0000 \rangle  = H_1  |0 \rangle \otimes H_1  |0 \rangle \otimes  H_1  |0 \rangle \otimes H_1  |0 \rangle=  
 \]
 \[
 \frac{1}{\sqrt{4^2}} \sum_{x \in B^4}  |x \rangle= \frac{1}{4}  \cdot ( |0000 \rangle +  |0001 \rangle +   |0010 \rangle +  |0011 \rangle +
 \]
 \[
 |0100 \rangle +  |0101 \rangle +   |0110 \rangle +  |0111 \rangle + |1000 \rangle +  |1001 \rangle +   |1010 \rangle +  |1011 \rangle +
\]
\[
( |1100 \rangle +  |1101 \rangle +   |1110 \rangle +  |1111 \rangle).
 \]
But how to generate the superposition $| \psi  \rangle$ of only four registers
\[
| \psi  \rangle= \frac{1}{2} \cdot \left(   |0011 \rangle + |1001 \rangle + |1111 \rangle +|0110 \rangle \right)
 \] 
with the resulting sparse amplitude representation $\alpha$?
\[
\alpha=
 \left( \begin{array}{r} 
0\\
0\\
0\\
\frac{1}{2}\\
0\\
0\\
\frac{1}{2}\\
0\\
0\\
\frac{1}{2}\\
0\\
0\\
0\\
0\\
0\\
\frac{1}{2}\\
 \end{array} \right)
\]
Basis encoding is the most intuitive method for encoding classical binary vectors into a quantum state. It transforms $m$-dimensional binary vectors into $m$-qubit quantum basis states.
 
 \subsection{Storage of Binary Vectors by Ventura and Martinez Algorithm}

 To generate a superposition of $k$ binary linear independent vectors with dimension $m$ with $n=2^m$ where $ k < n$, a method was proposed by Ventura and Martinez in 1988,  \cite{Ventura1988, Ventura2000} a and later simplified by Trugenberger, see \cite{Trugenberger2001}. The procedure involves successively dividing the current superposition into processing and memory branches. Each new memory branch is then populated with an input pattern step by step. The method is linear in the number of stored patterns and their dimension, \cite{Schuld2018}  with preparation cost $O(m \cdot k)$.
A related description with a step by step \textit{Qiskit}\footnote{ \textit{Qiskit} is an  open-source software development kit (SDK) for working with quantum computers at the level of circuits and~algorithms \cite{Qiskit2023},  IBM Quantum,  \textit{https:$//$quantum-computing.ibm.com$/$}.}   examples for storing three binary patterns with $m=2$ and $k=3$ can be found in \cite{Wichert2024}.
At the initial stage, the system is in its basis state, comprising load qubits, memory qubits, and control qubits denoted as $c_1$ and $c_2$.
 \[ | memory; c_2, c_1; load  \rangle
 \]
In our implementation, we utilize the \textit{Qiskit} little-endian notation. It is important to note that the original work employed big-endian notation, as indicated in \cite{Trugenberger2001}, $| register; c_1, c_2; memory  \rangle$.
The objective is to progressively decompose this basis state using the control register until the desired superposition  with the memory register in the required superposition
 \begin{equation}
\frac{1}{\sqrt{k}} \sum_{i=1}^k = | memory ;  0, 0;  0\cdots0   \rangle_i=  \left(\frac{1}{\sqrt{k}} \sum_{i=1}^k  | memory   \rangle_i \right)  \otimes  |0, 0;  0, \cdots, 0  \rangle. 
 \end{equation}
 We distinguish between a processing branch, indicated by the control qubit $c_2$ with the value one ($c_2=1$), and the memory branches, which are in superposition with the control qubit $c_2$ with the value zero ($c_2=0$).
The  control qubit  $c_1=1$  indicates  the split of the qubit $c_2$ by the operator $CS_p$ represented by the  parametrized  $CU$ gate $CU(\theta,\phi, \lambda)$ with  $ \phi=\pi, \lambda=\pi$, 
and $\theta=\arcsin\left(\frac{1}{\sqrt{p}}  \right) \cdot 2 $
 \[
CS_p= CU(\arcsin\left(\frac{1}{\sqrt{p}}  \right) \cdot 2,\pi, \pi)=\left( \begin{array}{cccc} 
 1 & 0 &  0 & 0 \\
 0 & 1 &  0 & 0 \\
 0 & 0 & \sqrt{\frac{p-1}{p}}  & \frac{1}{\sqrt{p}}   \\
 0 & 0 &\frac{-1}{\sqrt{p}}  & \sqrt{\frac{p-1}{p}}   \\
 \end{array} \right)
 \]
with $CS_p | c_2,c_1  \rangle$
 \[
CS_p | 01  \rangle = | 01  \rangle, ~~~~CS_p | 11  \rangle =  \frac{1}{\sqrt{p}} \cdot  |10\rangle +  \sqrt{\frac{p-1}{p}} \cdot |11 \rangle. 
 \]
 Since the control qubit  $c_1=1$ is entangled with the memory register
 we create the  memory branch ($c_2=0$) with  $\frac{1}{\sqrt{p}} \cdot | memory; 01  \rangle$ and processing  branch ($c_2=1$) $\sqrt{\frac{p-1}{p}} \cdot | memory; 11  \rangle$ by the split operation on the preceding processing  branch. We store the new pattern in the generated memory register of the new generated memory branch.

\subsubsection{Storage Algorithm}

The initial state is
\[
| \psi   \rangle_0 =  |  0\cdots0  ; 1, 0; pattern_1\rangle
\] 
with control qubit $c_2=1$ indicating that the state is a processing branch. The next steps describe a loop that
stores $k$ binary patterns
\[
| pattern_1 \rangle, | pattern_2 \rangle, \cdots | pattern_k \rangle
\]
\\
~\\
FOR $i=1$ TO $k$ 
\begin{itemize}
\item  We load the pattern  $pattern_i$ into the load register.
\item IF  $i==1$ THEN
\begin{itemize}
\item  We invert the ground state of the  memory register  of the processing branch to $|11\cdots1\rangle$  using  NOT gates. 
\end{itemize}
\item ELSE 
\begin{itemize}
 \item We copy the  $pattern_i$ into the memory register of the processing branch  $c_2=1$ using  ccX gate (CCNOT gate).
\item We copy the  $pattern_i$ into the memory register of the memory branch  $c_2=0$ and  processing branch  $c_2=1$  using  cX gate (CNOT gate).
 As a result  the memory register  of the processing branch is in the ground state $|00\cdots0\rangle$, this is not the case for the  memory register of memory  branches where the bits are flipped. 
 \item  We invert the ground state of the  memory register  of the processing branch to $|11\cdots1\rangle$  using  NOT gates to all memory registers.  Only the  memory register  of the processing branch represents  $|11\cdots1\rangle$. 
\end{itemize}
\item The control qubit  $c_1=1$  is entangled with the memory register $|11\cdots1\rangle$ by  the multi-controlled X gate.  As a result the control qubit  $c_1=1$  is entangled with the processing branch.
\item The processing branch is split  by the operator $CS_p$. with $p=m+1-i$.  Since the control qubit  $c_1=1$ is entangled with the memory register
 we create a new  memory branch  and a processing  branch. 
 \item We redo the entanglement of control qubit  $c_1=1$  with the memory register $|11\cdots1\rangle$.
 \item We redo the NOT gates operation to all memory registers.
\item We apply the CNOT gate controlled by the load register ($pattern_i$) to  the memory registers of all the branches.
As a result:
\begin{itemize}
\item The $pattern_i$ is represented in the new created memory registers of the memory branch and in the processing branch.
 \item The  memory registers of the already present memory branches are reconstructed by the flip back operation of the CNOT gate.
\end{itemize}
\item  We un-compute the memory register of the processing branch ($c_2=1$) to the ground state $|00\cdots0\rangle$ by the   ccX gate.
\item We reset the load register to the ground state
\end{itemize}
 NEXT\\
\begin{itemize}
\item We  convert the processing branch into a memory branch and store the last  pattern $| pattern_k \rangle$ in  its memory register.
\item We reset the load register to the ground state.
\item As a result we represent the stored patterns in the  superposition
 \[
| \psi   \rangle =  \left(\frac{1}{\sqrt{m}} \sum_{i=1}^k | pattern   \rangle_i  \right) \otimes  | 0, 0; 0\cdots0   \rangle.
 \]
 \end{itemize}

  \subsubsection{Example}
 
\begin{figure}[htb]
  \begin{center}
    \leavevmode
    \epsfysize5cm
    \epsffile{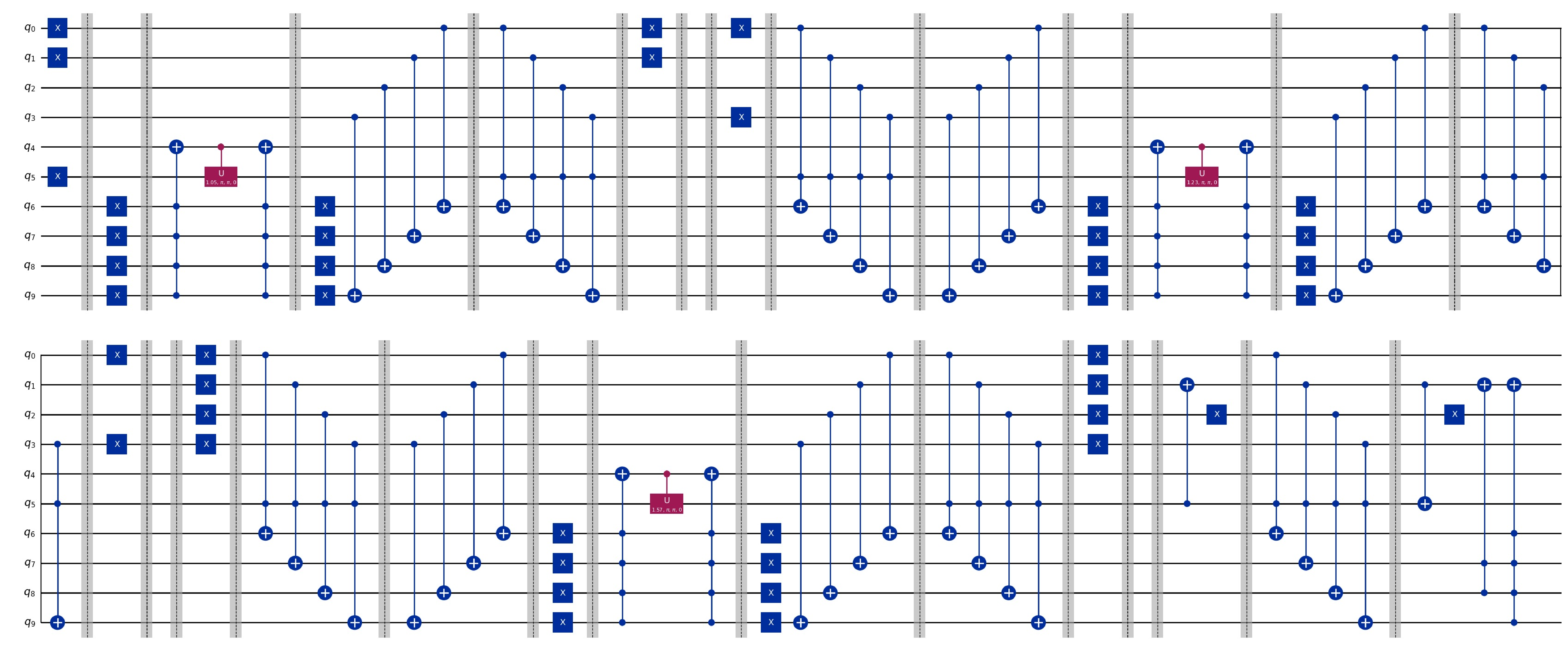}
  \end{center}
\caption{Circuit as proposed by generating the distribution $| \hat{\psi}  \rangle= \frac{1}{2} \cdot \left(   |0011 \rangle + |1001 \rangle + |1111 \rangle +|0110 \rangle \right) \otimes  | 000000   \rangle$.
 }
  \label{BasisEncod_4Pat.eps}
\end{figure}

 In this example we store four binary patterns,
 \[
  |0011 \rangle_1,~~  |1001 \rangle_2,~~ |1111 \rangle_3,~~ |0110 \rangle_4
\]
\textit{Qiskit}   uses little endian notation 
\[
| memory; c_2, c_1; register   \rangle = | q_9, q_8  q_7, q_6 ;q_5, q_4; q_3, q_2  q_1, q_0  \rangle.
\] 
Qubts $0$ - $3$ represent the  load register, qubits $4$ and $5$ are the control qubits  and the qubits $6$ - $9$ represent  the memory register. In Figure  \ref{BasisEncod_4Pat.eps} we represent the corresponding circuit that generates the required superposition \footnote{A  Jupyter notebook with a step by step description ising $Qiskit$ can be freely downloaded at https://github.com/andrzejwichert/Permutation-Trick.}
\[
| \hat{\psi}  \rangle= \psi \otimes  | 000000   \rangle = \frac{1}{2} \cdot \left(   |0011 \rangle + |1001 \rangle + |1111 \rangle +|0110 \rangle \right) \otimes  | 000000   \rangle.
 \] 
In our example, we require $30$ multi-controlled multi-not gates and $42$ controlled not gates, in addition to three controlled rotation gates.

 \subsection{Storage of Binary Vectors through Permutation Technique}
 
Given $k$ patterns represented by vectors of dimension $m$, where $k$ is a power of $2$ (i.e., $k = 2^g$), we can construct a superposition of $m$ qubits using a permutation operator.
To achieve this, we tensor the ground state of $m-g$ qubits with the uniform superposition of $g$ qubits. Subsequently, we map the resulting state using a permutation matrix $P$ into the desired superposition of $m$ qubits. This process generates the superposition $\psi$.
\[
| \psi  \rangle= \frac{1}{2} \cdot \left(   |0011 \rangle + |1001 \rangle + |1111 \rangle +|0110 \rangle \right).
 \] 
 with $k=4$ and $m=4$ by a tensor operation of the ground state  $( |00\rangle$ with uniform distribution of two qubits
 using a permutation operator  represented by a permutation matrix $P$   with 
 \[
 P \cdot (  |00 \rangle \otimes H  |0\rangle   \otimes H  |0 \rangle)=
 \]
 \[
 P \cdot  \frac{1}{2} \cdot ( |0000 \rangle + |0001  \rangle +  |0010 \rangle + |0011  \rangle) =
 \]
\begin{equation}
\frac{1}{2} \cdot ( 0011 \rangle +  |1001 \rangle + |1111 \rangle + |0110 \rangle.
\end{equation}
The application of the permutation operator through the matrix $P$ for quantum tree search was first introduced in \cite{Wichert2025}.
We employ little-endian notation and represent the mapping as
 \begin{equation}
\left( \begin{array}{r} 
0\\
0\\
0\\
\frac{1}{2}\\
0\\
0\\
\frac{1}{2}\\
0\\
0\\
\frac{1}{2}\\
0\\
0\\
0\\
0\\
0\\
\frac{1}{2}\\
 \end{array} \right)=
 P  \cdot
 \left( \begin{array}{r} 
\frac{1}{2}\\
\frac{1}{2}\\
\frac{1}{2}\\
\frac{1}{2}\\
0\\
0\\
0\\
0\\
0\\
0\\
0\\
0\\
0\\
0\\
0\\
0\\
 \end{array} \right).
\end{equation}
The matrix $P$  corresponds to  three permutations 
\[
  |0000 \rangle \leftrightarrow  |0110 \rangle 
\]
 \[
  |0001 \rangle \leftrightarrow  |1001 \rangle 
\]
 \[
  |0010 \rangle \leftrightarrow  |1111 \rangle 
\]
the last mapping is an equality
 \[
  |0011 \rangle =  |0011 \rangle. 
\]
Originating from the identity matrix of dimensions $16 \times 16$, the matrix $P$ is defined by the exchange of rows $1$ and $7$, rows $2$ and $10$, and rows $3$ and $16$. This operation corresponds to the addition of one to the represented binary values, as we assign the first row a value of one rather than zero. 
\[P=
\left(
\begin{array}{cccccccccccccccc}
 0 & 0 & 0 & 0 & 0 & 0 & 1 & 0 & 0 & 0 & 0 & 0 & 0 & 0 & 0 & 0 \\
 0 & 0 & 0 & 0 & 0 & 0 & 0 & 0 & 0 & 1 & 0 & 0 & 0 & 0 & 0 & 0 \\
 0 & 0 & 0 & 0 & 0 & 0 & 0 & 0 & 0 & 0 & 0 & 0 & 0 & 0 & 0 & 1 \\
 0 & 0 & 0 & 1 & 0 & 0 & 0 & 0 & 0 & 0 & 0 & 0 & 0 & 0 & 0 & 0 \\
 0 & 0 & 0 & 0 & 1 & 0 & 0 & 0 & 0 & 0 & 0 & 0 & 0 & 0 & 0 & 0 \\
 0 & 0 & 0 & 0 & 0 & 1 & 0 & 0 & 0 & 0 & 0 & 0 & 0 & 0 & 0 & 0 \\
 1 & 0 & 0 & 0 & 0 & 0 & 0 & 0 & 0 & 0 & 0 & 0 & 0 & 0 & 0 & 0 \\
 0 & 0 & 0 & 0 & 0 & 0 & 0 & 1 & 0 & 0 & 0 & 0 & 0 & 0 & 0 & 0 \\
 0 & 0 & 0 & 0 & 0 & 0 & 0 & 0 & 1 & 0 & 0 & 0 & 0 & 0 & 0 & 0 \\
 0 & 1 & 0 & 0 & 0 & 0 & 0 & 0 & 0 & 0 & 0 & 0 & 0 & 0 & 0 & 0 \\
 0 & 0 & 0 & 0 & 0 & 0 & 0 & 0 & 0 & 0 & 1 & 0 & 0 & 0 & 0 & 0 \\
 0 & 0 & 0 & 0 & 0 & 0 & 0 & 0 & 0 & 0 & 0 & 1 & 0 & 0 & 0 & 0 \\
 0 & 0 & 0 & 0 & 0 & 0 & 0 & 0 & 0 & 0 & 0 & 0 & 1 & 0 & 0 & 0 \\
 0 & 0 & 0 & 0 & 0 & 0 & 0 & 0 & 0 & 0 & 0 & 0 & 0 & 1 & 0 & 0 \\
 0 & 0 & 0 & 0 & 0 & 0 & 0 & 0 & 0 & 0 & 0 & 0 & 0 & 0 & 1 & 0 \\
 0 & 0 & 1 & 0 & 0 & 0 & 0 & 0 & 0 & 0 & 0 & 0 & 0 & 0 & 0 & 0 \\
 \end{array}
\right).
\]
In general, it is not possible to decompose a permutation matrix into a tensor product of other permutation matrices. This is because not all permutations can be expressed as a product of other permutations, and most permutations are not tensor products.
Instead, we will represent the mapping using quantum gates with preparation cost  $O(m \cdot k)$.

\subsubsection{General Storage Algorithm}

To store $k$ binary patterns 
\[
| pattern_1 \rangle, | pattern_2 \rangle, \cdots | pattern_k \rangle
\]
each represented by $m$ qubits,  we require an additional flag qubit. 
Consequently, we necessitate a total of $m+1$ qubits.
The following steps outline a loop that stores $k$ binary patterns:\\
~\\
FOR $i=1$ TO $k$ 
\begin{itemize}
\item Apply  Hadamard gate to the qubit $i-1$
\end{itemize}
NEXT i \\
FOR $i=1$ TO $k$ 
\begin{itemize}
\item  Entangle  the basis states $|0^{\otimes (m-k)} \rangle \otimes |i \rangle  $ with  the flag qubit 
\item  We write  $|pattern_i\rangle  $ into the superposition indicated by the flag qubit  by the controlled not gates 
\item We disentangle $|pattern_i\rangle  $ with the flag qubit indicated by the controlled not gates 
\end{itemize}
NEXT i\\
~\\
The complexity of the algorithm is $O(m \cdot k)$.

\subsubsection{Storage Algorithm Example}

We will use $4$ qubits to represent the superposition with an additional flag qubit. We generate uniform superposition of two qubits representing four states. 
We define the mapping between the ground state  $ |00\rangle$ with uniform distribution of two qubits
\[
 \frac{1}{2} \cdot ( |0000 \rangle + |0001  \rangle +  |0010 \rangle + |0011  \rangle) .
\]
For  the  uniform distribution  and distribution $| \psi \rangle$
we define a mapping between by each basis state  with
\[
  |0000 \rangle \rightarrow  |0110 \rangle 
\]
 \[
  |0001 \rangle \rightarrow  |1001 \rangle 
\]
 \[
  |0010 \rangle \rightarrow  |1111 \rangle 
\]
 \[
  |0011 \rangle =  |0011 \rangle 
\]
as before. 
In our circuit, we generate a uniform distribution on two qubits by employing two Hadamard gates.
Next, we address each basis state, indicating the flag qubit $4$, using a multi-controlled NOT gate. We then write the required mapping using flag-controlled NOT gates and disentangle the flag qubit using a multi-controlled NOT gate.
It is important to note that we do not need to write $|0011\rangle = |0011\rangle$ since it is already present.
We require $2 \cdot 3$ multi-controlled NOT gates and $6$ controlled NOT gates. 
 \begin{figure}[htb]
  \begin{center}
    \leavevmode
    \epsfysize2cm
    \epsffile{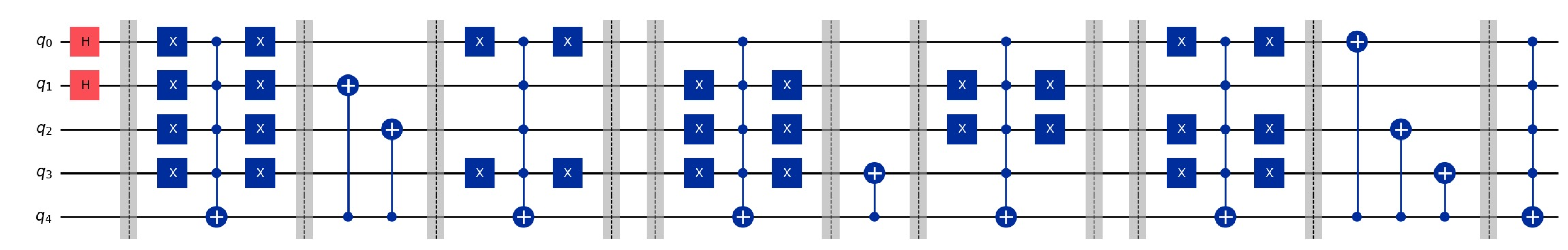}
  \end{center}
\caption{Proposed  circuit generating the distribution $| \psi  \rangle=   \frac{1}{2} \cdot \left(   |0011 \rangle + |1001 \rangle + |1111 \rangle +|0110 \rangle \right)$. We define the  ground state  $|00\rangle$  by the qubits $2$ and $3$ with uniform distribution of two qubits by two Hadamard gates, qubits $0$ and $1$.
Then we address each basis states indicating the flag qubit $4$ by multi controlled not gate, write the required mapping with flag controlled not gates and disentangle the flag qubit with multi controlled not gates gate. Note we do not need to write $ |0011 \rangle =  |0011 \rangle$ since it already present.
We require $2 \cdot 3$ multi controlled  multi controlled not gates and $6$ controlled not gates.
 }
  \label{Encod_4Pat.eps}
\end{figure}

\subsubsection{Reduce Algorithm}

We minimize the number of gates by mapping the basis states in a uniform distribution efficiently and writing equal $1$ qubits in patterns in parallel.
\begin{itemize}
\item We compute the Hamming distance $d_H$ between the basis states in the uniform distribution and the patterns that we intend to store.
We represent this distance as a distance matrix $D$.
\item WHILE matrix $D$ not EMPTY
\begin{itemize}
\item We identify the minimal Hamming distance(s) and determine the corresponding mappings
 \item We remove the corresponding columns and rows from the matrix $D$
\end{itemize}
\item We compute the similarity between the patterns that we intend to store. This similarity is represented by the similarity matrix $S$
 \item Identify clusters of patterns and their mappings with high similarity values in the matrix $S$
\item  Load the mapping of clusters in parallel through entanglement into superposition
\end{itemize}

\subsubsection{Reduce Algorithm Example}

\begin{figure}[htb]
  \begin{center}
    \leavevmode
    \epsfysize2cm
    \epsffile{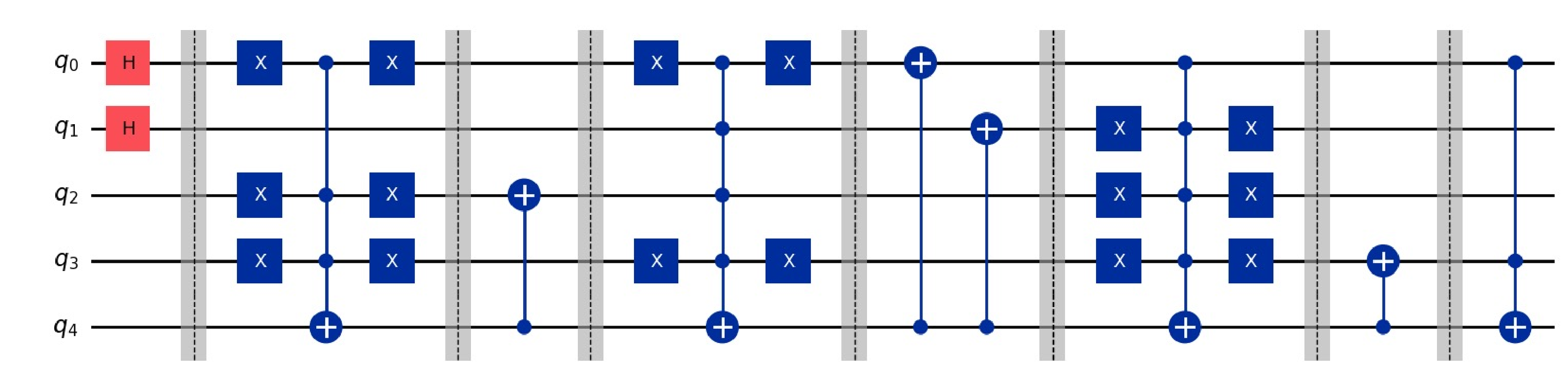}
  \end{center}
\caption{Proposed  circuit generating the distribution $| \psi  \rangle$ by parallel load as determined by the $reduce$ algorithm.}
  \label{Compress_Parallel_A.eps}
\end{figure}

\begin{figure}[htb]
  \begin{center}
    \leavevmode
    \epsfysize2cm
    \epsffile{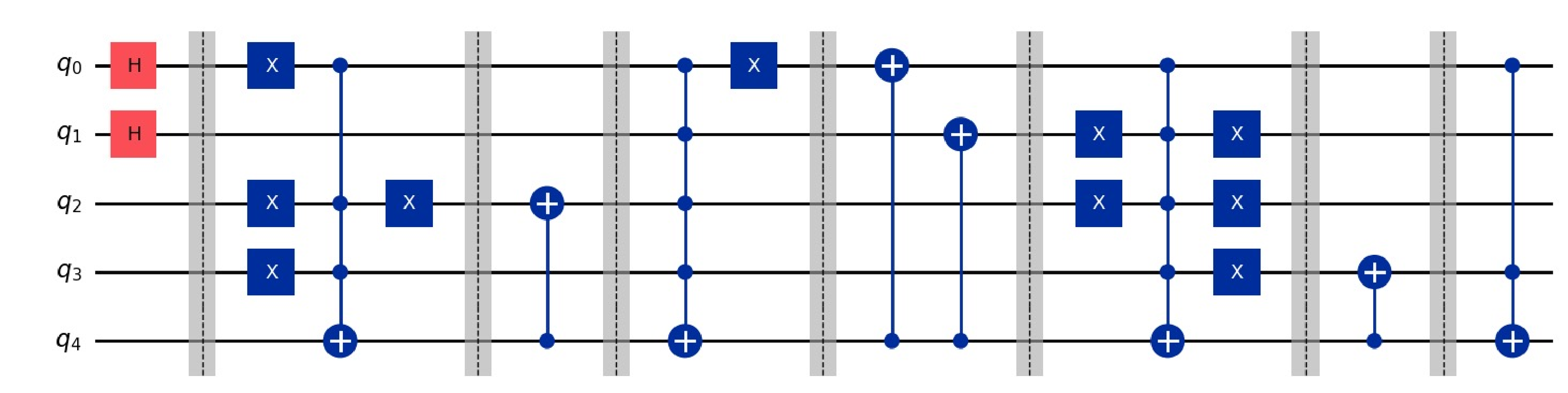}
  \end{center}
\caption{Proposed circuit generating the distribution $| \psi \rangle$ by parallel load as determined by the $reduce$ algorithm without unnecessary NOT gates.} 
  \label{Compress_Parallel.eps}
\end{figure}

The algorithm reduces the number of gates by mapping the basis states in the uniform distribution efficiently and writing equal $1$ qubits in patterns in parallel.
\paragraph{Efficient mapping :}
To ascertain the optimal mapping, we calculate the Hamming distance $d_H$ between the basis states in the uniform distribution and the patterns to be stored. This distance is then represented in a distance matrix:
\[
\begin{array}{c|cccc}
 d_H   & |0011 \rangle  & |0110 \rangle  & |1001 \rangle  &|1111 \rangle \\ \hline
 |0000 \rangle  & 2  & 2 & 2 & 4 \\
  |0001 \rangle  & 1 & 2   & 1 & 3 \\
  |0010 \rangle  & 1 & 1 & 2   & 3 \\
 |0011 \rangle & \textbf{0} & 1 & 1 & 2
\end{array}
\]
We identify the minimal Hamming distance(s) of zero that corresponds to the identity mapping:
\[
|0011 \rangle = |0011 \rangle.
\]
We eliminate the column and row representing the state $|0011\rangle$, which corresponds to a Hamming distance of zero. This operation yields the following distance matrix:
\[
\begin{array}{c|ccc}
 d_H      & |0110 \rangle  & |1001 \rangle  &|1111 \rangle \\ \hline
 |0000 \rangle   & 2 & 2 & 4 \\
  |0001 \rangle   & 2   &  \textbf{1} & 3 \\
  |0010 \rangle   &  \textbf{1}  & 2   & 3 \\
\end{array}
\]
We identify the minimal Hamming distances of $1$, which corresponds to two mappings;
\[
 |0001 \rangle  \rightarrow  |100\textbf{1} \rangle 
\]
\[
 |0010 \rangle \rightarrow   |01\textbf{1}0 \rangle 
\]
We eliminate the columns and rows that have Hamming distances of 1. The remaining mapping is
\[
 |0000 \rangle \rightarrow   |1111 \rangle. 
\]
\paragraph{Parallel writing of equal $1$ qubits: } 
Utilizing the efficient mapping, we proceed to write the remaining equal $1$ qubits in parallel patterns. This is indicated by the similarity matrix.
Qubits that have already been written are denoted by $I$ and are not considered when determining the equal $1$ qubits:
\[
\begin{array}{c|ccc}
  sim  & |01I0 \rangle  & |100I \rangle  &|1111 \rangle \\ \hline
   |01I0 \rangle   & *  & 0 & 1 \\
|100I \rangle  & 0 & *   & 1\\
|1111 \rangle \rangle & 1 & 1 & *
\end{array}
\]
We can concurrently write the following two clusters, with cluster $I$ writing qubit $2$ in parallel\footnote{\textit{Qiskit} uses little-endian notation with qubits indexes $|q_3 q_2 q_1 q_0  \rangle$},
\[
 \left( 0010 \rangle \rightarrow  \right)~  |0\textbf{1}I0 \rangle
\]
  \[
\left( 0000 \rangle \rightarrow  \right)~ |1\textbf{1}11 \rangle
\]
and  cluster $II$ writing qubit  $3$ in parallel
\[
 \left( 0001 \rangle \rightarrow  \right)~  |\textbf{1}00I \rangle 
\]
 \[
\left( 0000 \rangle \rightarrow  \right)~ |\textbf{1}111 \rangle. 
\]
The clusters are identified within the algorithm through the concept of entanglement.
\paragraph{Algorithm :}
We entangle the flag qubit with the states $|0000\rangle$ and $|0010\rangle$ (cluster II), as indicated by $E$.
\[
 |0000\rangle~E,~~|0010 \rangle~E   
 \]
and write in parallel qubit in qubit $2$
\[
  |0000 \rangle ~E \rightarrow  |0\textbf{1}00 \rangle~E
\]
 \[
  |0010 \rangle ~E \rightarrow  |0\textbf{1}10 \rangle ~E
\]
 We disentangle the state $|0110 \rangle$ indicated by $D$
 \[
|0100 \rangle~E,~~|0110 \rangle~D
\]
We  write in the entangled state $ |0100 \rangle$  qubits $0$ and $1$
 \[
  |0100 \rangle~E   \rightarrow  |01\textbf{1} \textbf{1} \rangle ~E
\]
  We entangle the state $|0001 \rangle$ (cluster $I$) with two entangled states
 \[
 |0111\rangle~E,~~|0001 \rangle~E   
 \]
 and write in parallel qubit in qubit $3$
\[
  |0111\rangle \rightarrow   |\textbf{1}111 \rangle ~E 
\]
 \[
  |0001 \rangle \rightarrow   |\textbf{1}001 \rangle ~E
\]
Finally, we disentangle the two resulting states.
\[
|\textbf1111 \rangle ~D,~~  |\textbf1001 \rangle ~D
\]
resulting in the distribution $| \psi \rangle$ as shown in the Figure  \ref{Compress_Parallel_A.eps}. We necessitate four multi-controlled NOT gates and four controlled NOT gates, which represents a substantial reduction. We can further minimize the number of unnecessary NOT gates, resulting in the circuit depicted in Figure \ref{Compress_Parallel.eps}.

 \subsection{Comparison of Both Methods}
 
 Both methods exhibit linear complexity in terms of the number of stored patterns and their dimensionality, resulting in a complexity of $O(m \cdot k)$.
The Ventura and Martinez (VM) algorithm necessitates $2 \cdot m + 2$ qubits, whereas the permutation technique (PT) requires $m + 1$ qubits, which is half the number of qubits required by the VM algorithm.
However, the permutation technique (PT) imposes a constraint on the number of stored patterns $k$, which must be a power of 2, denoted as $k = 2^g$. 
 In the following Table \ref{VMPT} we indicate number of some complex gates $H$, $U$ (rotation gate),  $CCX$ and $MCX$ gate.
 It should be noted that the number of  $MCX$ gates in the Permutation Technique (PT) can be reduced considerably through the $reduce$ algorithm. 
\begin{table} [h!]
\caption{We indicate number of some complex gates $H$, $U$ (rotation gate),  $CCX$ and $MCX$ gate.}
%\begin{center}
\begin{tabular} {| c | c | c | c | c |} \hline
 &  H & U &  CCX &  MCX \\ 
\hline \hline
VM 	& -	&  $k-1$ &  $k \cdot m$ & $(k-1) \cdot 2$ \\ \hline
PT 	& $k$   	&  -		&  - 	& $k \cdot 2$  \\ \hline
\end{tabular}
\label{VMPT}
%\end{center}
\end{table}

 \section{Retrieval of Binary Vectors}

Retrieval of Binary Vectors is mostly based Grover's algorithm in which the result is marked by an oracle.
An alternative method of quantum counting  based on Euler’s formula  was proposed by \cite{Trugenberger2001}.

 Grover's amplification \index{Grover, Lov K.} algorithm implements  exhaustive search in $O( \sqrt{n})$ steps in  $n$-dimensional Hilbert   \cite{grover1996},  \cite{grover1997},  \cite{grover1998a},  \cite{grover1998b},  \cite{grover1996},  \cite{grover1996}. It is as good as any possible quantum algorithm for exhaustive search \index{exhaustive search} due to the \index{lower bound} lower bound $\Omega( \sqrt{n})$ \cite{aharonov1999}. The algorithm is based on the Householder reflection of state $| x \rangle$ of $m$ qubits   \cite{zalka1999a}.
For a oracle  $o(x)$
\begin{equation}
  o_{\xi}(x)= \left\{
  \begin{array}{l} 
1~~~ if~~~x=\xi  \\
 0~~~else
 \end{array}  \right.
 \end{equation}
 we seek to identify the value of $x$ representing a pattern  with  $o(x)=1$, where $x=\xi$.
 
Grover’s amplification algorithm implements exhaustive search in $O(\sqrt{n})$ steps in an $n$-dimensional Hilbert space   \cite{grover1996},  \cite{grover1997},  \cite{grover1998a},  \cite{grover1998b},  \cite{grover1996},  \cite{grover1996}. 
This algorithm is derived from the Householder reflection of the state $|x\rangle$ of $m$ qubits, where $n=2^m$. Grover’s amplification algorithm is optimal for exhaustive search, as demonstrated by the lower bound $\Omega(\sqrt{n})$ established by  \cite{aharonov1999}.
Grover's amplification algorithm is optimal, one can prove that a better algorithm cannot exist \cite{bennett1997},  \cite{boyer1998}. 
Uniform distributions are crucial for the operation of Grover’s algorithms. If the distribution is non-uniform, the algorithms may not function correctly or necessitate adaptation, resulting in the same complexity of $O(\sqrt{n})$. For instance, an adapted algorithm for sparse distributions, where the majority of amplitudes are zero, must generate a uniform distribution for the unmarked sets, leading to the same complexity of $O(\sqrt{n})$.

 \subsection{Ventura Martinez Trick}

In the quantum associative memory (QuAM), as proposed by Venture and Martinez, a modified version of Grover’s search algorithm is employed to determine the answer vector corresponding to a query vector with the  $O(\sqrt{n})$ Grover's rotation. This approach is compared to the conventional computer’s search algorithm, which  $O(k)$ search costs. The QuAM algorithm offers an advantage if the following condition is met:
\[
\sqrt{n} < k.
\] 
To understand the modified version of Grover’s search algorithm we generate simple superposition with $k=4$ and $n=16$, see as well \cite{Wichert2024},
\[
| \psi  \rangle= \frac{1}{2} \cdot \left(   |0011 \rangle + |1001 \rangle + |1111 \rangle +|0110 \rangle \right).
 \] 
 so that we can track the amplitude distribution 
\[
\frac{1}{2} (0,0,0,1,0,0,1,0,0,1,0,0,0,0,0,1)^T.
\]
We mark the target state $|0110 \rangle$ by a negative phase
\[
\frac{1}{2} (0,0,0,1,0,0,-1,0,0,1,0,0,0,0,0,1)^T.
\]
and perform a Grover's rotation with the result
\[
\frac{1}{8} (-1,-1,-1,3,-1,-1,-5,-1,-1,3,-1,-1,-1,-1,-1,3)^T
\]
We repeat the process and mark the target state $|0110 \rangle$ by applying a negative phase. Subsequently, we perform a Grover’s rotation on the resulting state
\[
 (-0.2, -0.2, -0.2, 0.3,-0.2,-0.2,0.6,-0.2,-0.2,0.3,-0.2,-0.2,-0.2,-0.2,-0.2,0.3)^T.
\]
The four states that represent our distribution exhibit high non-negative amplitudes (the results are rounded for clarity).
After marking the target state $|0110 \rangle$ by a negative phase and a Grover’s rotation, 
\[
 (0.02, 0.02, 0.02, 0.5,0.02,0.02,-0.4,0.02,0.02,0.5,0.02,0.02,0.02,0.02,0.02,0.5)^T
\]
all four basis states have nearly equal distribution, the information of the target state $|0110 \rangle$ is negative. Marking the target state with a negative phase will result in the loss of information about it.

\begin{figure}[htb]
  \begin{center}
    \leavevmode
    \epsfysize3.5cm
    \epsffile{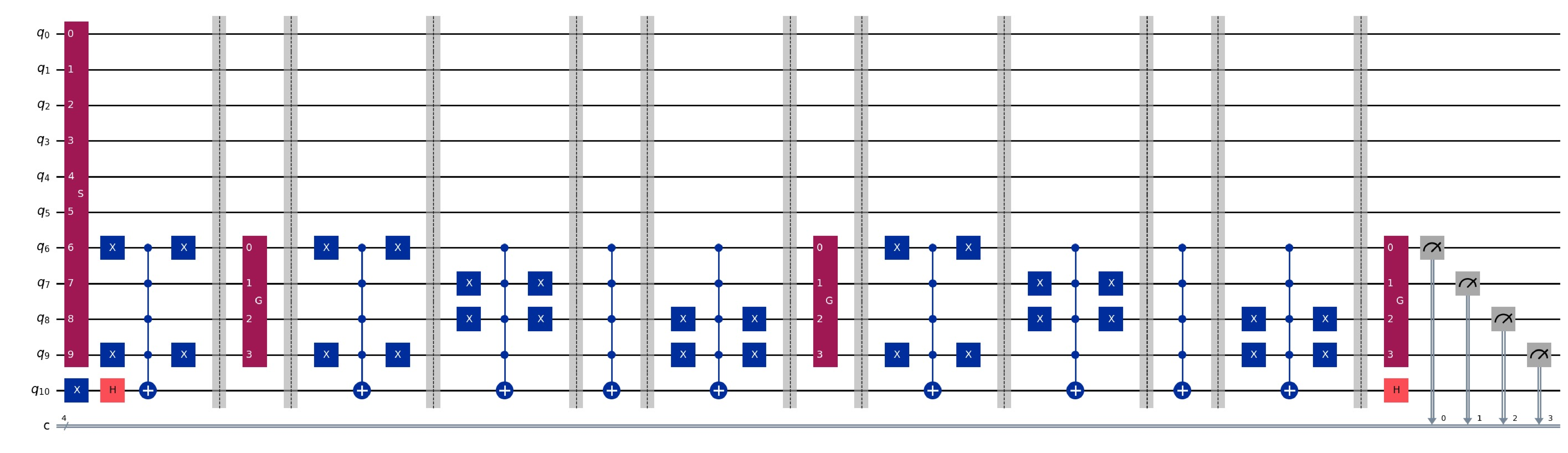}
  \end{center}
 \caption{The circuit representing the distribution $\psi$ indicated by the gate $S$ followed by three Grover’s rotations (gate $G$) with twice the cost of marking all four stored patterns, indicating the target pattern $|0110\rangle$.}
 \label{Grover_Asso_VM.eps}
\end{figure}

Venture and Martinez, proposed a modified version of Grover’s search algorithm \cite{Ventura1988, Ventura2000}.  
As previously established,
From the initial distribution, we mark the target state $|0110 \rangle$ by a negative phase.
\[
\frac{1}{2} (0,0,0,1,0,0,-1,0,0,1,0,0,0,0,0,1)^T.
\]
and perform a Grover’s rotation with the result
\[
\frac{1}{8} (-1,-1,-1,3,-1,-1,-5,-1,-1,3,-1,-1,-1,-1,-1,3).^T
\]
However, currently, all four patterns that represent our distribution are marked by a negative phase, and a Grover’s rotation is performed with the result
\[
\frac{1}{8} (1,1,1,-1,1,1,7,1,1,-1,1,1,1,1,1,-1)^T
\]
We repeat the process of marking all four patterns that represent our distribution by a negative phase and then perform a Grover’s rotation
\[
 (0,0,0,0,0,0,1,0,0,0,0,0,0,0,0,0)^T
\]
with the amplitude one indicating the target pattern $|0110 \rangle$, we require three Grover’s rotations, which entail a cost equivalent to marking all four present states. This is illustrated in Figure \ref{Grover_Asso_VM.eps}, which depicts the circuit for our example. 
Following the initial Grover’s amplification, we must mark all $k$ stored patterns in the subsequent rotation. The number of rotations is independent of $k$ and solely dependent on $n$, where $n = 2^m$. Consequently, we require $O(\sqrt{n})$ rotations, compared to $O(k)$ queries on a conventional computer.
For four patterns in superposition, our preceding example and one target value, we would only need one rotation. This improvement can be achieved through the permutation trick.

\subsection{Retrieval  by Permutation Trick}

 We generate the distribution $| \psi  \rangle$ by applying the operator $P$ (as depicted in Figure \ref{Compress_Parallel.eps}, which illustrates $P$ using a compressed circuit). Subsequently, we record the outcome of the query vector by the oracle,
\[
| \psi  \rangle=  \frac{1}{2} \cdot \left(   |0011 \rangle + |1001 \rangle + |1111 \rangle - |0110 \rangle \right).
\]
We map  the distribution $| \psi  \rangle$
with the marked pattern (or patterns) by the circuit $P^T$ into the distribution $|\gamma \rangle$
\begin{equation}
P^T: | \psi  \rangle \rightarrow |00\rangle \otimes  | \gamma  \rangle
\end{equation}
\begin{equation}
| \gamma  \rangle=    \frac{1}{2} \cdot ( |11  \rangle + |01  \rangle+  |00 \rangle  - |10 \rangle ). 
 \end{equation}
in  four dimensional Hilbert space $\mathcal{H}_4$.
Applying a single rotation of Grover’s algorithm to the distribution $| \gamma \rangle$ results in the distribution $| \zeta \rangle$, which is represented by the state $|10 \rangle$. We subsequently map the distribution $| \zeta \rangle$ into the distribution $|\Psi \rangle$ using the following mapping function:
 \begin{equation}
 P:  |00 \rangle \otimes   | \zeta  \rangle \rightarrow   |\Psi \rangle
\end{equation}
  \begin{equation}
 P: |00 \rangle \otimes  |10 \rangle \rightarrow  |0110 \rangle
\end{equation}
 and measure the distribution $|\Psi \rangle$ with the result $|0110 \rangle$.
 
 The initial distribution $|\psi  \rangle$ is represented in $n=2^m$ dimensional Hilbert space $\mathcal{H}_n$ by $P$.
  We mark the queries by an oracle in  $\mathcal{H}_n$.
 We map $|\psi \rangle$ into a decomposed space  $\mathcal{H}_n =\mathcal{H}_z \otimes \mathcal{H}_k$ with $k=2^g$ and  $z=2^{(m-g)}$.
 In the Hilbert space $\mathcal{H}_z$ we represent the ground state and  in  $\mathcal{H}_k$ the distribution $| \gamma \rangle$.
 We preform Grover's rotation in the   $\mathcal{H}_k$ and map the resulting distribution  into   $\mathcal{H}_n$. After  $\tau=\sqrt{k}$ rotations  we measure the result of $|\Psi ^\tau \rangle$ in the Hilbert space $\mathcal{H}_n$.
 Each rotation is composed of marking of the queries by an oracle in  $\mathcal{H}_n$, mapping into a decomposed space  $\mathcal{H}_n =\mathcal{H}_z \otimes \mathcal{H}_k$ by $P^T$, Grover's rotation in the   $\mathcal{H}_k$ and mapping into Hilbert space $\mathcal{H}_n$ by $P$.
\begin{figure}[htb]
  \begin{center}
    \leavevmode
    \epsfysize4cm
    \epsffile{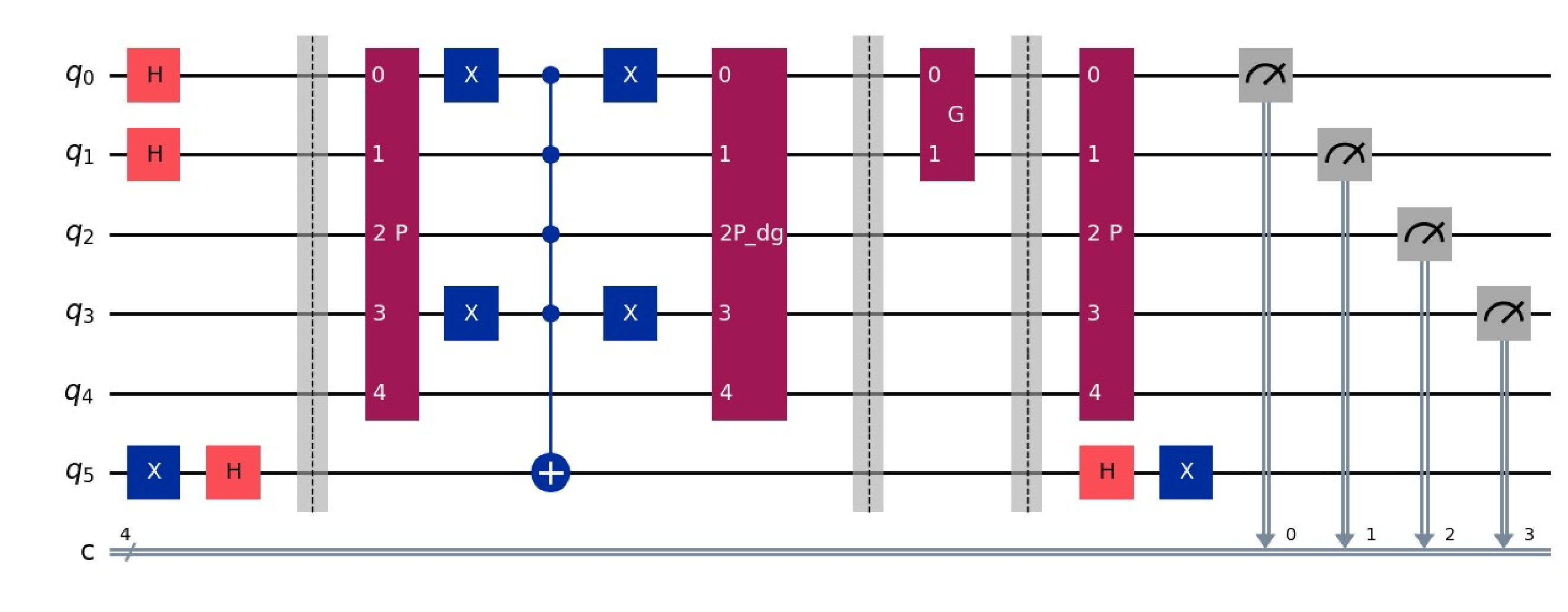}
  \end{center}
 \caption{The initial distribution $|\psi\rangle$ is represented in a 16-dimensional Hilbert space $\mathcal{H}_{16}$ by $P$.
We mark the query by an oracle in $\mathcal{H}_{16}$.
We map $|\psi\rangle$ into a decomposed space $\mathcal{H}_{n=16} = \mathcal{H}_{z=4} \otimes \mathcal{H}_{k=4}$.
In the Hilbert space $\mathcal{H}_{z=4}$, we represent the ground state, and in $\mathcal{H}_{k=4}$, we represent the distribution $| \gamma \rangle$ by by $P^T$.
We perform one Grover’s rotation in the $\mathcal{H}_{k=4}$ and map the resulting distribution into $\mathcal{H}_{n=16}$ by $P$. We measure the result in the Hilbert space $\mathcal{H}_{n=16}$.}
  \label{Grover_Asso_Parallel..eps}
\end{figure} 
In the quantum associative memory  (QuAM)  as proposed by Venture and Martinez the number of rotations of the Grover's algorithm is  independent of $k$ with  $O(\sqrt{n})$ rotations. Using the permutation trick the  number of rotations is independent  of $n$ with $O(\sqrt{k})$ rotations.

 \subsection{Comparison of Both Methods}
 
The QuAM algorithm, employing the Ventura Martinez technique, exhibits a superior performance compared to conventional computer search algorithms under the following condition:
\[
\sqrt{n} < k.
\]
Given that we are required to mark $k$ stored patterns of dimension $m$, the cost would be
\[
O(m \cdot k \cdot \sqrt{n})
\]
which is substantially higher compared to the corresponding cost on conventional computers, which is $O(m \cdot k)$. This disparity arises due to the overhead incurred as a result of the necessity of marking all $k$ stored patterns after the initial Grover’s rotation. 

To implement a permutation trick, we must mark $k$ stored patterns of dimension $m$ twice. This involves applying the permutation function $P$ and its transpose $P^T$ to these patterns. The cost associated with this process is given by the expression:
\[
O(m \cdot k \cdot \sqrt{k})
\]
The primary advantage of the proposed approach is its dense coding, which necessitates $m$ qubits to store $k$ patterns. In Table \ref{VMPT2}, we compare both methods.
\begin{table} [h!]
\caption{We indicate number of Grover's rotations and oracle calls.}
%\begin{center}
\begin{tabular} {| c | c | c | } \hline
 &  Grover's Rotations & Oracle Calls \\  \hline \hline
VM 	& $ \sqrt{n}$ &  $\sqrt{n-1} \cdot k+1$  \\ \hline
PT 	&   $\sqrt{k}$  & 	$\sqrt{k} \cdot k \cdot 2$  \\ \hline
\end{tabular}
\label{VMPT2}
%\end{center}
\end{table}
 It is noteworthy that the number of $MCX$ gates representing oracle calls in the permutation technique (PT) can be significantly reduced through the $reduce$ algorithm. If the number of oracle calls $p$ is less than $\sqrt{k}$, we gain an advantage over conventional search because
\[
\sqrt{k} \cdot p < k,
\]
which implies a quadratic reduction in oracle calls through parallelization during the storage phase. This reduction can be achieved if the data exhibits high correlation.

\section{Discussion}

Quantum machine learning algorithms, including quantum associative memory, face the input destruction problem, see \cite{Aimeur2013,Wittek14, Aaronson2015}. This problem arises because the storage phase incurs preparation costs, and the measurement phase involves a collapse, rendering the information created in the storage phase unusable. This bottleneck for data encoding necessitates efficient data preparation, although it is often overlooked.
Theoretical speedups are analyzed in the context of efficient data preparation. For highly correlated data, such as when the $k$ binary vectors exhibit a quadratic correlation, the permutation technique, implemented through the $reduce$ algorithm, can be employed to achieve efficient data preparation.

\section{Conclusion}

We introduced a novel quantum storage algorithm for $k$ binary vectors of dimension $m$ with $k < 2^m$ into a superposition of an $m$-qubit quantum state using a permutation technique. Additionally, we introduced a \textit{reduce} algorithm that can lead to further reduction through parallel operations.
We described a modified version of Grover’s algorithm based on the permutation trick, which enables an exhaustive search by Grover’s algorithm in $O(\sqrt{k})$ steps for $k$ patterns, independent of $n = 2^m$. We compared this algorithm to the Ventura Martinez trick, which requires $O(\sqrt{n})$ steps for $k$ patterns. 

\backmatter

\bmhead{Supplementary information}

All $Qiskit$ examples are presented in a Jupyter Notebooks that can be freely downloaded at https://github.com/andrzejwichert/Permutation-Trick.

\section*{Compliance with Ethical Standards}

The funders had no role in study design, data collection and analysis, decision to publish, or preparation of the manuscript. The authors declare no conflicts of interest. This article does not contain any studies with human participants or animals performed by any of the authors.

%% BioMed_Central_Bib_Style_v1.01

\end{document}